\documentclass[aps,superscriptaddress,prapplied,reprint,twoside]{revtex4-2}
\usepackage[utf8]{inputenc}
\usepackage{enumitem}
\usepackage{float}
\usepackage{graphicx}
\usepackage{dcolumn}
\usepackage{bm}
\usepackage{hyperref}
\usepackage{amsmath}
\usepackage{xcolor}
\begin{document}
\raggedbottom

\title{FID Magnetometer Based on Paraffin-Coated Planar Reflective Multipass Cells}
\author{Xiangyu Li}
\author{Tingxuan Guo}
\author{Yang Li}
\author{Kaifeng Zhao}
\email{zhaokf@fudan.edu.cn}
\affiliation{Key Laboratory of Nuclear Physics and Ion-Beam Application (MOE), Fudan University, Shanghai 200433, China}
\affiliation{Institute of Modern Physics, Department of Nuclear Science and Technology, Fudan University, China} 
\date{\today}

\begin{abstract}
We demonstrate a paraffin-coated planar reflective multipass vapor cell for compact optical atomic magnetometry. The cell has an internal volume of $12 \times 12 \times 8~\mathrm{mm}^3$ and supports 20 optical passes with a total transmittance exceeding $65\%$, while maintaining a longitudinal spin-relaxation time of $^{87}\mathrm{Rb}$ longer than $1~\mathrm{s}$. The planar geometry provides spatially separated input and output beams, enabling compact optical integration. A single-cell free-induction-decay (FID) magnetometer reaches $10~\mathrm{pT}/\sqrt{\mathrm{Hz}}$ in the geomagnetic-field range, presently limited by current-source noise in the field coils. A two-cell differential configuration achieves a sensitivity of $\sim 28~\mathrm{fT}/\sqrt{\mathrm{Hz}}$ over the $1$--$15~\mathrm{Hz}$ band for bias fields of $0.3$--$0.7~\mathrm{G}$. These results establish that paraffin-coated planar multipass cells offer high optical depth, long coherence times, and an integration-friendly platform for ultrasensitive magnetometry.
\end{abstract}

\maketitle
\section{\label{sec1}Introduction}
Optical atomic magnetometers (OAMs)~\cite{Budker2007} are among the most sensitive platforms for detection of ultra-weak to geomagnetic magnetic fields, combining high field sensitivity with low power consumption, compactness, and the potential for portable operation. These attributes have enabled applications in human biomagnetism~\cite{Boto2018,Brookes2022,Kominis2003}, archaeology~\cite{Fassbinder2023,Becker1995}, particle physics~\cite{Safronova2018,Brown2010,Feng2022}, mineral exploration~\cite{Lu2023}, and aerospace sensing~\cite{Korth2016,Bennett2021}. For many of these applications, however, practical deployment requires not only high intrinsic sensitivity, but also reduced sensor volume, low operating temperature, mechanical robustness, and compatibility with integrated optical architectures. Multipass atomic vapor cells directly address these requirements by increasing the effective atom--light interaction length within a compact physical volume, thereby enhancing optical rotation signals and improving magnetometric sensitivity without proportionally increasing the sensor size~\cite{Sheng2013,Lucivero2021,Liu2022,Cai2020,Yu2022}. 

Conventional multipass vapor cells are most commonly based on Herriott-cell geometries~\cite{Herriott1965,Li2011} or related variants~\cite{Cooper2022}, in which a laser beam undergoes repeated reflections between two curved mirrors. In typical implementations, one mirror contains a centered through-hole that serves as both the input and output port for the laser beam, and the vapor cell is fabricated using anodic bonding technology~\cite{Wallis1969,Kanda1990,Knowles2006}. Although Herriott-type cells provide long optical paths in compact volumes, their curved internal reflective surfaces impose several constraints. Repeated beam focusing can increase sensitivity to atomic diffusion-induced relaxation~\cite{Li2011,Sheng2013}, while the through-hole geometry complicates optical alignment and limits the spatial separation between incident and exiting beams. An alternative approach uses parallel planar mirrors mounted outside the vapor cell~\cite{Li2022a,Qi2025,Patent2023}. In such planar multipass configurations, the beam propagates without the frequent focusing inherent to curved-mirror cavities, which can reduce diffusion-related relaxation and simplify beam routing. The spatial separation of the input and output beams also provides a natural advantage for miniaturization and system integration. However, because the mirrors are externally mounted, it incurs additional optical loss from repeated transmission through the vapor-cell windows and is susceptible to mechanical vibrations.

A second major challenge arises from spin relaxation in small-volume vapor cells. As the cell volume decreases, polarized atoms collide more frequently with the cell walls, making relaxation induced by wall collisions the dominant limitation on the atomic coherence time. Many multipass vapor cells therefore employ buffer gases to suppress wall relaxation~\cite{Romalis1997,Li2021}. However, this approach introduces collisional broadening and requires a higher operating temperature, which increases spin-exchange relaxation and limits the achievable sensitivity~\cite{Oreto2004}. An alternative approach is to apply an anti-relaxation coating to the inner surface of the vapor cell. In a multipass configuration, optical losses accumulate exponentially with the number of reflections, requiring the coating to be highly transparent to minimize its impact on the overall transmission. OTS (octadecyltrichlorosilane) self-assembled coatings exhibit high optical transparency due to their minimal thickness of only a few molecular layers, but their anti-relaxation performance remains limited~\cite{Seltzer2009,Zhang2015,Li2022b}. Paraffin coatings provide superior anti-relaxation performance~\cite{Bouchiat1966,Graf2005,Balabas2010}; however, their thickness and uniformity are harder to control. Excessive deposition on the reflective surfaces can degrade optical reflectivity and induce beam divergence. Furthermore, the presence of internal mirrors increases the geometric complexity of the inner volume, making it more challenging to form a uniform and transparent paraffin coating in Herriott-type multipass cells.

Here we introduce a planar reflective multipass vapor cell that integrates high-reflection dielectric films directly onto two parallel inner walls of the vapor cell. This architecture combines optical-path enhancement with a simple geometry and mechanical robustness. It differs from a conventional vapor cell only through the addition of dielectric reflective films, providing a simple surface geometry that is favorable for uniform anti-relaxation coating deposition. We demonstrate that a paraffin anti-relaxation coating is compatible with repeated multipass reflection: the coating allows up to $4\times10^4$ wall collisions without depolarization while introducing limited additional optical power loss. The enhanced optical path increases the signal amplitude sufficiently to enable operation at relatively low temperature and with small-angle rf excitation, which suppresses nonlinear Zeeman effects and improves the robustness of frequency extraction. Using two such multipass vapor cells, we realize an atomic free-induction-decay magnetometer with a differential sensitivity of $\sim 28~\mathrm{fT}/\sqrt{\mathrm{Hz}}$ under geomagnetic-scale fields.

\section{\label{sec2}Fabrication and Characterization of Paraffin coated Multipass Vapor Cells}

\subsection{Fabrication of Planar Reflective Multipass Cells}
\begin{figure}
  \centering
  \includegraphics[width=1\columnwidth]{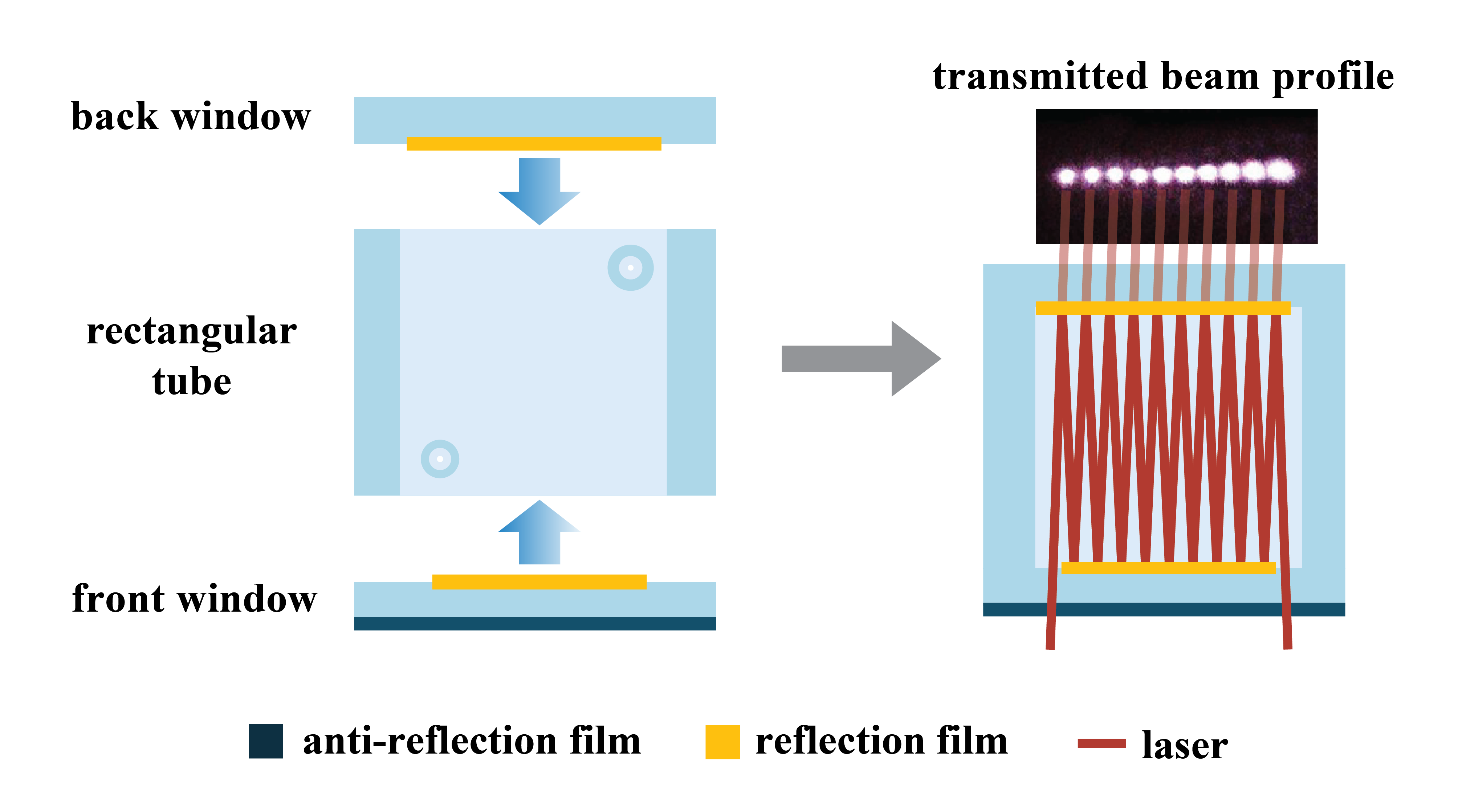}
  \caption{%
Fabrication and optical characterization of the planar reflective multipass vapor cell. Left: Schematic illustration of the coated-window assembly joined by optical contact bonding. The two diagonally positioned circles indicate the cell stems. Right: Multipass beam geometry. Upper right: Infrared image of the transmitted beam spots with ten reflections from the back window (corresponding to 20 optical passes). The spot size increases for later passes because of beam divergence during propagation.
  }
  \label{fig:bonding}
\end{figure}

An overview of the cell fabrication process is shown in Fig.~\ref{fig:bonding}. A rectangular borosilicate glass tube (Yixing Jingke Optical Instrument) is prefabricated with two stems on the same side, one long and one short. Each stem is connected to the tube interior through a $0.2$--$0.3~\mathrm{mm}$ laser-drilled pinhole in the side wall. Two borosilicate glass plates, serving as the front and back windows of the cell, are prepared with dimensions matching the end faces of the tube. The tube ends and window surfaces are precision polished to a flatness better than $\lambda/10$ at $633~\mathrm{nm}$ and a surface quality of $20/10$. The inner surface of each window is coated with a dielectric reflective film consisting of alternating $\mathrm{SiO}_2$ and $\mathrm{Ta}_2\mathrm{O}_5$ with $\mathrm{SiO}_2$ as the top layer, yielding a reflectivity exceeding $99.5\%$. This reflective film is confined to the central region of the window, leaving an uncoated margin for bonding to the tube. On the front window, additional uncoated regions are reserved as optical entrance and exit ports. The outer surface of the front window is coated with a dielectric antireflection film. The two windows are then prebonded to the rectangular tube by optical contact bonding~\cite{Haisma2002}. This method allows direct, adhesive-free joining of glass components at room temperature through intermolecular forces. The assembled cell is subsequently annealed at $600~^\circ\mathrm{C}$ in an electric furnace to form a permanent, vacuum-tight bond.

The multipass performance of the cell is characterized by sending a $780~\mathrm{nm}$ diode laser into the entrance port of the front window. The number of passes is determined by imaging the transmitted spot pattern on a screen placed behind the back window with an infrared camera. Each transmitted spot corresponds to one reflection from the back window, as shown in Fig.~\ref{fig:bonding}; therefore, the total number of light passes is twice the number of observed spots. The pass number can be tuned by adjusting the incident angle. For cells with a $12 \times 8~\mathrm{mm}^2$ cross section and a length of $12~\mathrm{mm}$, a 20-pass configuration with $84\%$ transmission is achieved. The cells used in the following experiments all have this dimensions and geometry.

\subsection{Paraffin Coated Multipass Vapor Cells}
To suppress wall relaxation of spin polarization, the inner surfaces of the planar-multipass cell are coated with a paraffin-based anti-relaxation coating. Tetratetracontane ($\mathrm{C_{44}H_{90}}$, Merck), which has a melting point of $85~^\circ\mathrm{C}$, is used as the coating material. The coating procedure follows a method similar to that described in Ref.~\cite{Alexandrov2002}.

The cell interior is first cleaned with $15\%$ hydrochloric acid and then thoroughly rinsed with deionized water. The cleaning solution is introduced through the long stem, while the short stem acts as a vent for air release. This two-stem geometry facilitates efficient solution exchange and minimizes liquid trapping inside the cell. After cleaning, the cell is dried in a vacuum oven.

The short stem is then flame-sealed, and the cell is connected through the long stem to a glass manifold attached to a vacuum system~\cite{Singh1972}. A small glass-sealed iron rod with a small amount of paraffin adhered to its surface is preloaded into the manifold. The manifold section containing the cell but excluding the iron rod is heated with a detachable oven at $400~^\circ\mathrm{C}$ for $12~\mathrm{h}$ under continuous pumping. After cooling to room temperature, the system pressure reaches several $10^{-7}~\mathrm{Torr}$.

The paraffin carrier is guided through the manifold by an external magnet. The paraffin is then melted with a hot-air gun and transferred into the stem of each cell. The stems are flame-sealed to detach the individual cells from the manifold, trapping a small amount of paraffin inside each cell. A final evaporative coating step is performed by heating each cell in a commercial convection oven with a cavity size of $45 \times 45 \times 45~\mathrm{cm}^3$ at $380~^\circ\mathrm{C}$ for $1~\mathrm{h}$ and allowing it to cool naturally. To reduce temperature gradients that could degrade coating uniformity, a hole is drilled at the center of one sidewall of the oven, allowing an externally mounted motor to rotate an internal rack holding the cell at a constant speed of $5~\mathrm{rpm}$ during coating.

After each evaporative coating step, the cell is inspected under an optical microscope for excessive paraffin deposition on the reflective surfaces, and its multipass performance is evaluated. In this work, cells with a 20-pass transmittance below 50\% are returned to the convection oven for recoating, before which the cell position is readjusted so that the window with excessive paraffin deposition is held at a slightly higher temperature than in the previous trial. Typically, one to three coating trials are required to satisfy the transmittance criterion.

\begin{figure}
  \centering
  \includegraphics[width=0.49\columnwidth]{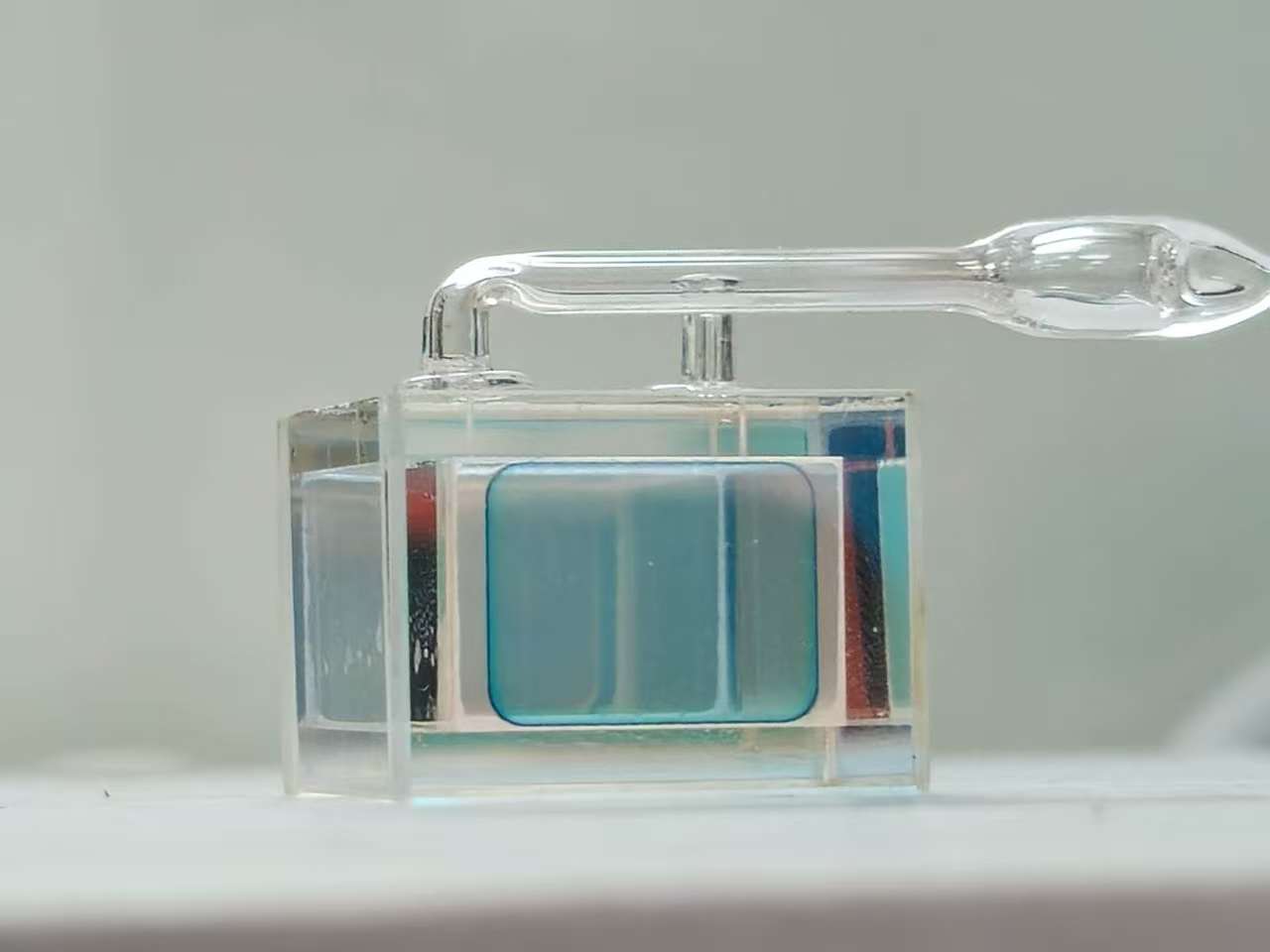}
   \put(-115,80){\textcolor{white}{\textbf{(a)}}}
  \hfill
  \includegraphics[width=0.49\columnwidth]{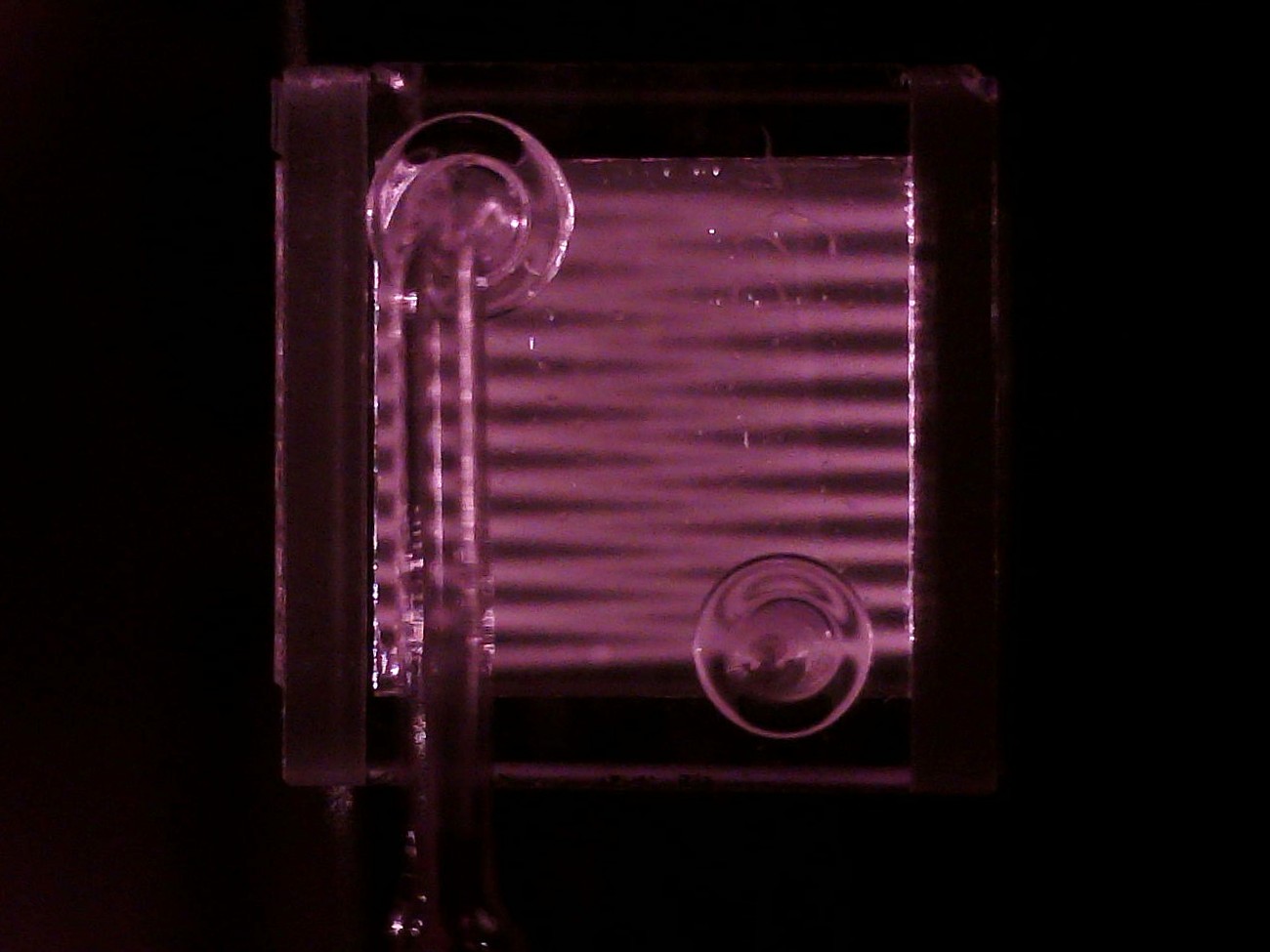}
   \put(-115,80){\textcolor{white}{\textbf{(b)}}}
  \caption{Photographs and optical characterization of paraffin-coated planar multipass vapor cells. (a) Photograph of a completed cell with internal dimensions of $12~\mathrm{mm}\times 8~\mathrm{mm}$(cross section)$\times 12~\mathrm{mm}$(length) and a wall thickness of $2~\mathrm{mm}$. (b) Infrared image showing resonant fluorescence along the optical beam path inside the cell, corresponding to 19 reflections and 20 light passes.
  }
  \label{fig:mp_cell}
\end{figure}

The paraffin-coated cell is subsequently reconnected to the vacuum system for rubidium filling, after which the $^{87}\mathrm{Rb}$ vapor cells are cured at $85~^\circ\mathrm{C}$ for several days in a laboratory water bath.  A completed vapor cell exhibits a transmittance of $65\%$ to $75\%$ for 20 optical passes. Fig.~\ref{fig:mp_cell}(a) shows a photograph of a representative cell. Fig.~\ref{fig:mp_cell}(b) shows an infrared fluorescence image obtained with a resonant probe beam, directly visualizing the multipass optical trajectory inside the cell.

\begin{figure}
  \centering
  \includegraphics[width=\columnwidth]{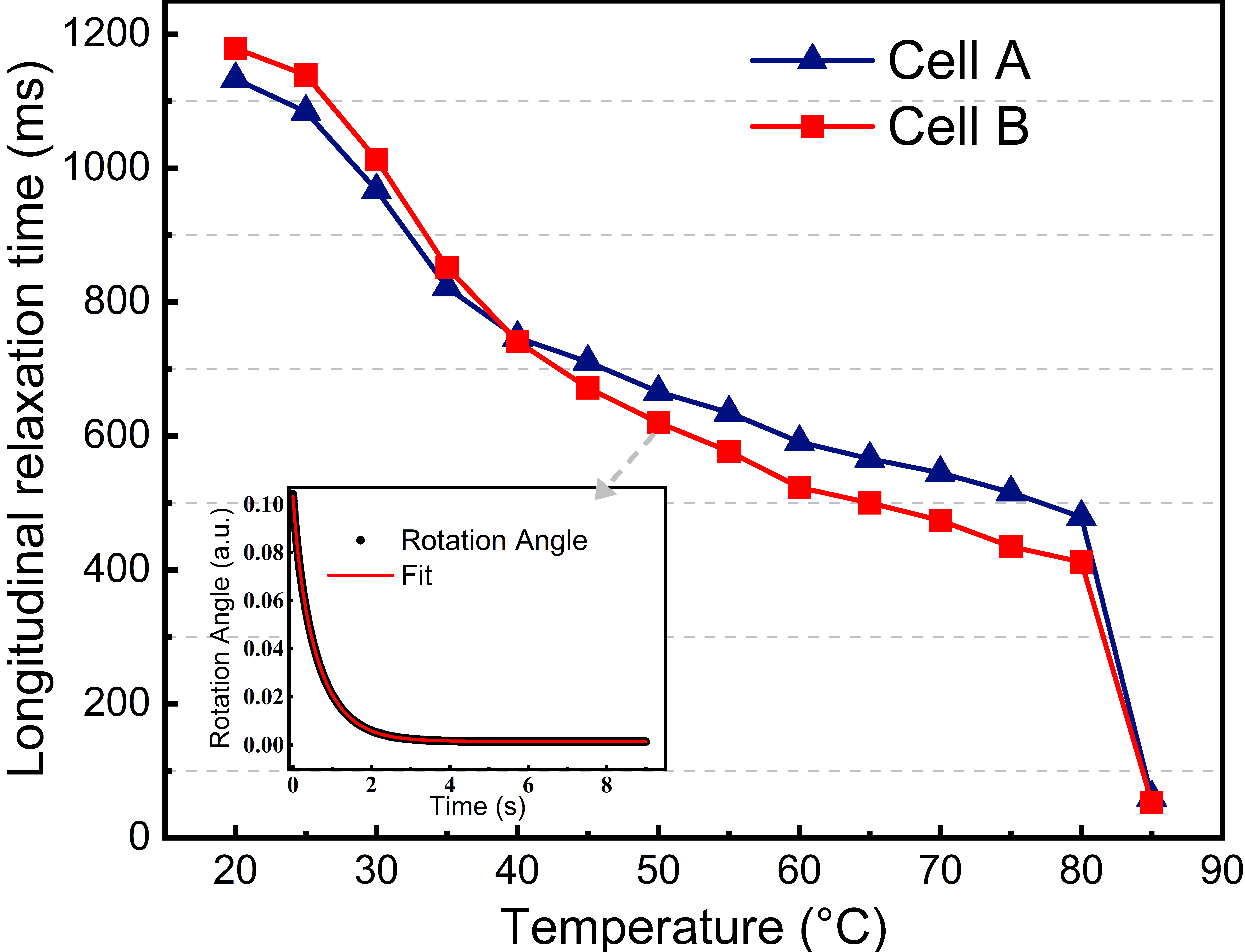}
  \caption{
 Temperature dependence of the longitudinal relaxation time $T_1$ for two paraffin-coated planar multipass vapor cells. The inset shows an exponential fit to the longitudinal spin relaxation of cell B at $50~^\circ\mathrm{C}$.
 }
  \label{fig:T1}
\end{figure}

The longitudinal spin-relaxation time $T_1$ is measured using a dark-relaxation method. A holding magnetic field of $0.2~\mathrm{G}$ is generated by three Helmholtz coils without magnetic shielding. A $5~\mathrm{mW}$ circularly polarized pump beam, resonant with the $F=1 \rightarrow F'=2$ transition of the $^{87}\mathrm{Rb}$ $D_1$ line, propagates along the magnetic-field direction and prepares the atomic ensemble in a spin-oriented state in the $F=2$ hyperfine level. The incident probe beam, linearly polarized with a power of $50~\mu\mathrm{W}$ and red-detuned by approximately $5~\mathrm{GHz}$ from the $F=2 \rightarrow F'=1$ transition of the $^{87}\mathrm{Rb}$ $D_1$ line, enters the cell through the uncoated window and propagates in a counterpropagating configuration to monitor the optical rotation induced by the spin orientation. The pump beam is gated by a mechanical optical shutter (SR470, Stanford Research Systems). After the pump beam is switched off, the optical-rotation signal exhibits a biexponential decay consisting of a minor fast component and a dominant slow component. The time constant of the slow component is extracted and identified as $T_1$.

Fig.~\ref{fig:T1} shows the temperature dependence of $T_1$ for the two vapor cells used in this work, denoted as Cell A and Cell B, both having same dimensions as that of Fig.~\ref{fig:mp_cell}. At room temperature, both cells exhibit $T_1$ values exceeding $1.1~\mathrm{s}$, corresponding to approximately $4.4 \times 10^4$ wall collisions without depolarization. With increasing temperature, $T_1$ decreases gradually and reaches approximately $0.6~\mathrm{s}$ at $50~^\circ\mathrm{C}$. A more pronounced reduction occurs above $80~^\circ\mathrm{C}$, suggesting the onset of paraffin melting.

\section{Magnetometry Application}
\begin{figure}
  \centering
  \includegraphics[width=\columnwidth]{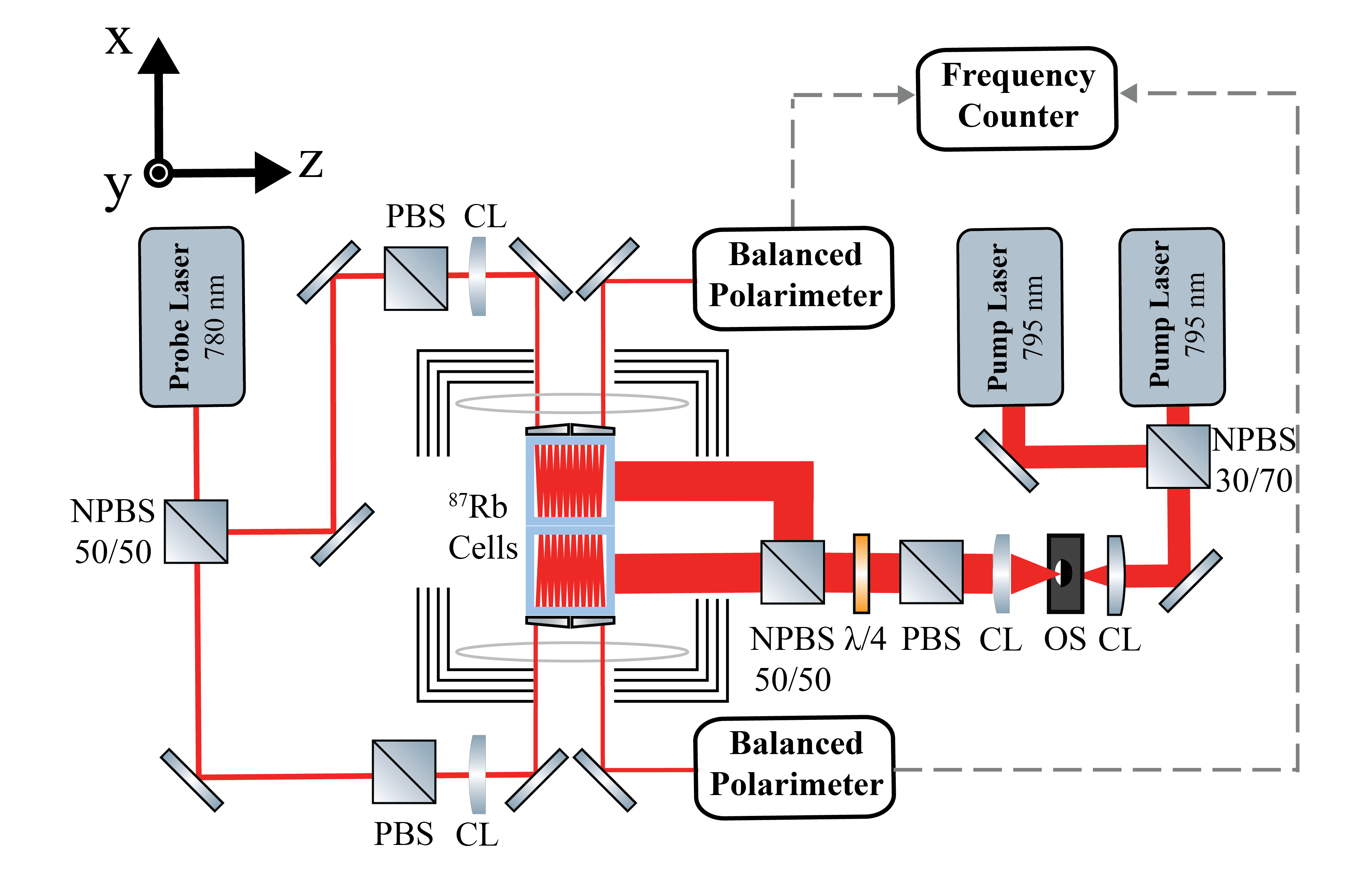}
  \caption{%
  Schematic of the experimental setup. NPBS: non-polarizing beam splitter; CL: convex Lens; PBS: polarizing beam splitter; OS: optical shutter; $\lambda$/4: quarter-wave plate; $\lambda$/2: half-wave plate. A Barker coil system on the inner cylinder of the magnetic shield (not shown) generates a static magnetic field along the $z$ axis. A separate pair of Helmholtz coils produces an rf magnetic field along the $x$ axis. 
  }
  \label{fig:experiment_setup}
\end{figure}

To demonstrate the capability of the paraffin-coated multipass cells for magnetometry, we implement a magnetic gradiometer based on two such cells, as illustrated in Fig.~\ref{fig:experiment_setup}. A four-layer cylindrical $\mu$-metal shield is used to suppress environmental magnetic-field noise. A Barker coil system (Shanghai Jieling Magnetic Material \& Devices) generates a geomagnetic-scale magnetic field along the $z$ axis, while a second pair of Helmholtz coils produces an oscillating rf magnetic field along the $x$ axis. Two vapor cells, Cell A and B as shown in Fig.~\ref{fig:T1}, are mounted side by side along the $x$ axis at the center of the coils. The cells are enclosed in a boron nitride oven, which is heated by a $10~\mathrm{MHz}$ ac-driven electric heating film and stabilized at $50~^\circ\mathrm{C}$.

Two $795~\mathrm{nm}$ lasers, tuned to the $F=1 \rightarrow F'=2$ and $F=2 \rightarrow F'=2$ transitions of the $^{87}\mathrm{Rb}$ $D_1$ line, are combined to form the pump beam. The combined beam is circularly polarized and gated by a mechanical optical shutter. A 50:50 nonpolarizing beam splitter (NPBS) divides the pump beam into two beams, one for each vapor cell. Each pump beam propagates along the $z$ axis, with the $F=1 \rightarrow F'=2$ and $F=2 \rightarrow F'=2$ components having powers of $2.0~\mathrm{mW}$ and $0.3~\mathrm{mW}$, respectively. The resulting spin polarization is measured to exceed 99\%~\cite{Julsgaard2003a}.

A separate $795~\mathrm{nm}$ laser, red detuned by approximately $10~\mathrm{GHz}$ from the $F=2 \rightarrow F'=1$ transition of the $^{87}\mathrm{Rb}$ $D_1$ line, is split by a 50:50 NPBS into two $200~\mu\mathrm{W}$ probe beams, one for each vapor cell. Each probe beam is linearly polarized along the $z$ axis and propagates along the $x$ axis. A pair of antireflection-coated wedges mounted at the optical slits of the cells minimizes the angle between the incident and exiting beams, allowing unobstructed optical access through the apertures of the magnetic shield. The optical-rotation signal induced by spin polarization in each cell is detected by a polarimeter consisting of a polarizing beam splitter and a balanced photodetector. The signals from the two cells are recorded simultaneously by the two channels of a Keysight 53220A frequency counter.

\begin{figure}[H]
  \centering
  \includegraphics[width=1\columnwidth]{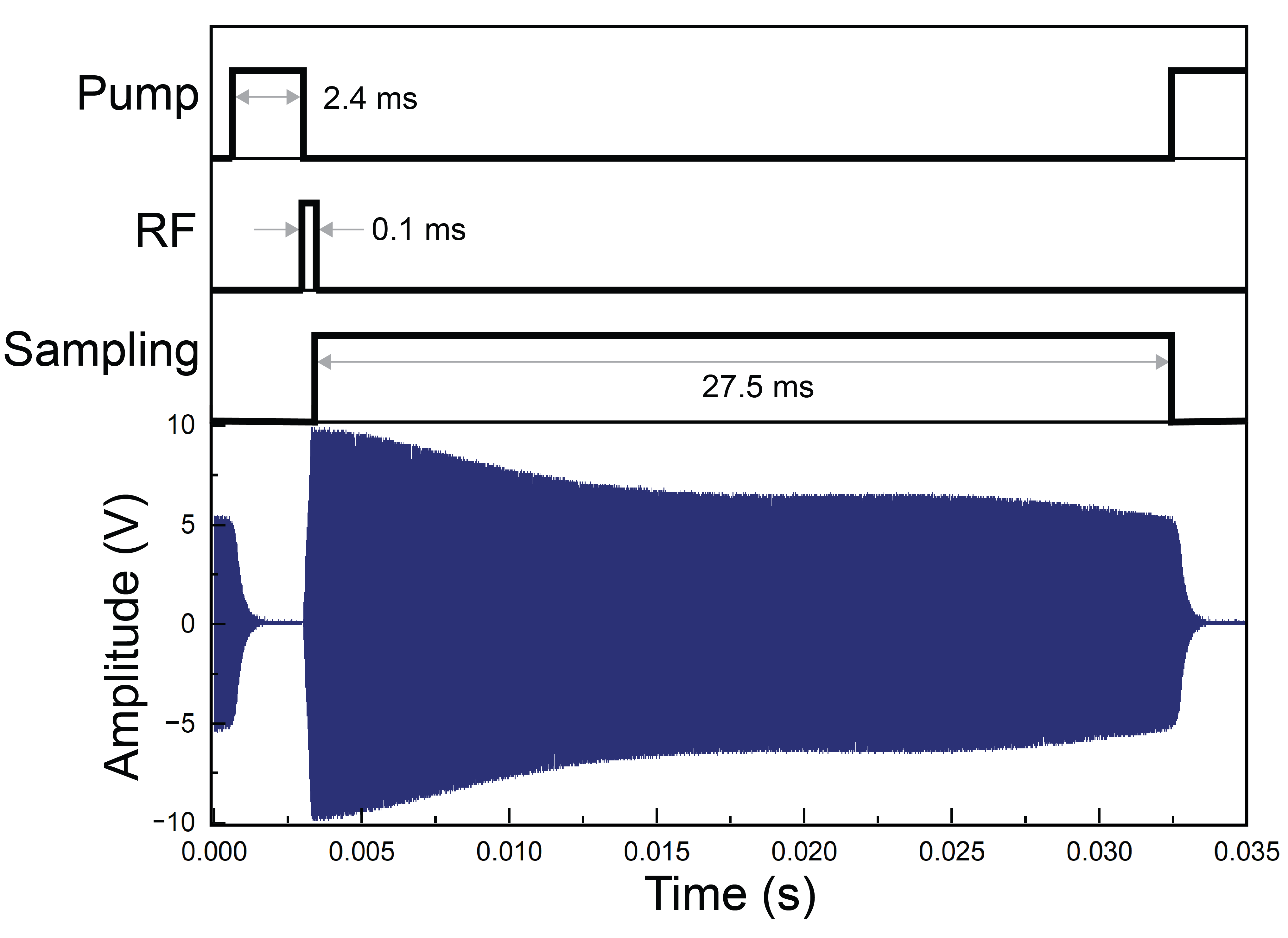}
  \caption{%
  Control timing sequence. The pump duration, rf pulse duration, and frequency-counter gate time are $2.4~\mathrm{ms}$, $0.1~\mathrm{ms}$, and $27.5~\mathrm{ms}$, respectively, resulting in a total measurement cycle of 33~ms. The actual repetition period is 3~ms longer than the nominal sum of the pulse durations. This difference arises because only part of the FID signal is used for counter-based frequency extraction, and finite delays are inserted between pulses to ensure proper timing synchronization.
  }
  \label{fig:sequence}
\end{figure}

The control timing sequence is illustrated in Fig.~\ref{fig:sequence}. The sequence begins with a $2.4~\mathrm{ms}$ optical-pumping pulse, which prepares the atomic ensemble in a stretched state. The pulse duration is limited by the closing time of the mechanical shutter. A $0.1~\mathrm{ms}$ near-resonant rf pulse is then applied to tilt the atomic spin polarization toward the $xy$ plane by an angle $\theta = \gamma B_{\mathrm{rf}}\tau/2$,
where $\gamma$ is the gyromagnetic ratio, $B_{\mathrm{rf}}$ is the amplitude of the rf magnetic field, and $\tau$ is the rf-pulse duration. After the rf excitation, the frequency counter is triggered to measure the free-induction-decay frequency of the transverse spin polarization during a $27.5~\mathrm{ms}$ gate time. The sequence is then repeated.

The decaying free precession of the spin can be described as a superposition of damped oscillatory components:
\begin{equation}
V_{\mathrm{sig}}(t) \propto
\sum_{m=-1}^{2}A_m\cos\left(\omega_m t + \phi_m\right)\exp\left(-\Gamma_m t\right),
\label{eq:FID_signal2}
\end{equation}

where $\omega_m = (E_m - E_{m-1})/\hbar$ is the frequency of the component associated with the splitting between adjacent Zeeman levels, and $\Gamma_m$, $\phi_m$, and $A_m(\theta)$ are the transverse relaxation rate, initial phase, and relative amplitude of the corresponding component, respectively. 

If the atoms are prepared in a stretched state prior to rf excitation, the relative amplitudes of the four frequency components are determined solely by the tilt angle $\theta$ of the collective spin polarization and follow a power-law dependence on $\tan(\theta/2)$:
\begin{equation}
A_2:A_1 : A_0 : A_{-1}=
1 :
3\tan^2\left[\frac{\theta}{2}\right] :
3\tan^4\left[\frac{\theta}{2}\right] :
\tan^6\left[\frac{\theta}{2}\right].
\label{eq:Am_ratio}
\end{equation}

\begin{figure}[t]
  \centering
  \includegraphics[width=1\columnwidth]{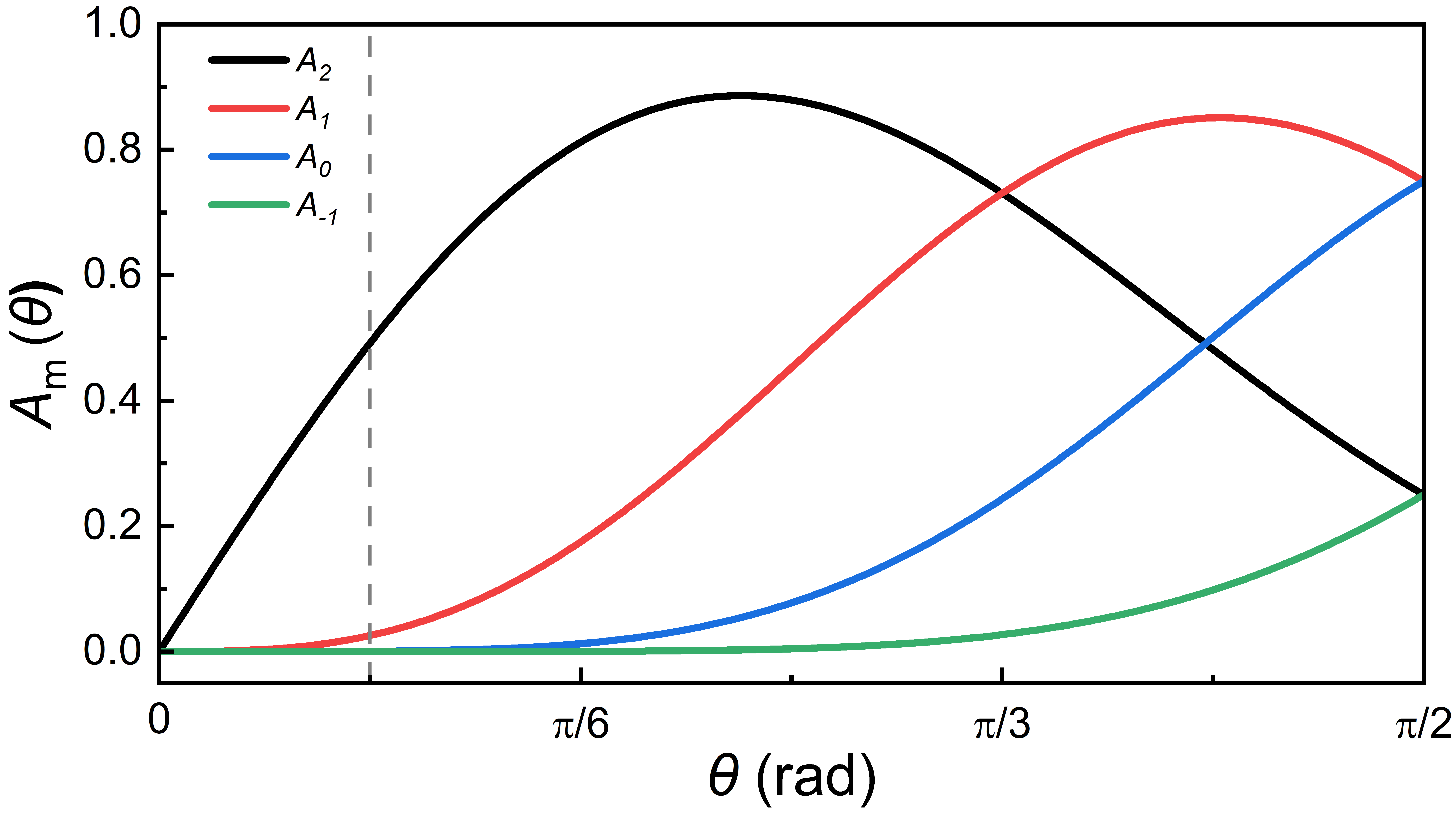}
  \caption{%
  Amplitudes $A_m(\theta)$ as functions of the spin tilt angle $\theta$. In the small-angle regime, lower-$m$ frequency components are suppressed, and the FID spectrum is dominated by the $A_2$ component.
  }
  \label{fig:Am}
\end{figure}

Fig.~\ref{fig:Am} shows the dependence of the amplitudes of the frequency components on the tilt angle.
For geomagnetic fields, the nonlinear Zeeman effect causes the splitting of  $\omega_m $ comparable or greater than $\Gamma_m$. A large tilt angle increases transverse polarization but also populates multiple Zeeman components with different precession frequencies, leading to faster dephasing of $V_{\mathrm{sig}}(t)$. In the small-angle regime ($\theta \ll 1$), lower-$m$ components are strongly suppressed, and the FID spectrum is dominated by the $A_2$ term, suppressing the nonlinear Zeeman effect and extending the coherence time. In our experiment, a tilt angle of $\theta = \pi/12$ is used to ensure operation in the small-angle regime. The enhanced signal provided by the multipass cell allows a small-tilt-angle rf pulse to generate an FID signal with sufficient signal-to-noise ratio. Therefore, the anti-relaxation capability paraffin coating can be fully exploited, extending the effective coherence time no less than the measurement window of $27.5~\mathrm{ms}$ and enhancing the magnetic-field sensitivity. 

The FID signals from the two channels are converted into discrete frequency time series using a frequency counter operated in gated mode. The frequency data are then analyzed by fast Fourier transform (FFT) to obtain the noise power spectral density and the corresponding magnetic-field sensitivity. For each channel, the measured sensitivity is $10~\mathrm{pT}/\sqrt{\mathrm{Hz}}$, mainly limited by noise from the current source that generates the bias magnetic field.

\begin{figure}[t]
  \centering
  \includegraphics[width=1\columnwidth]{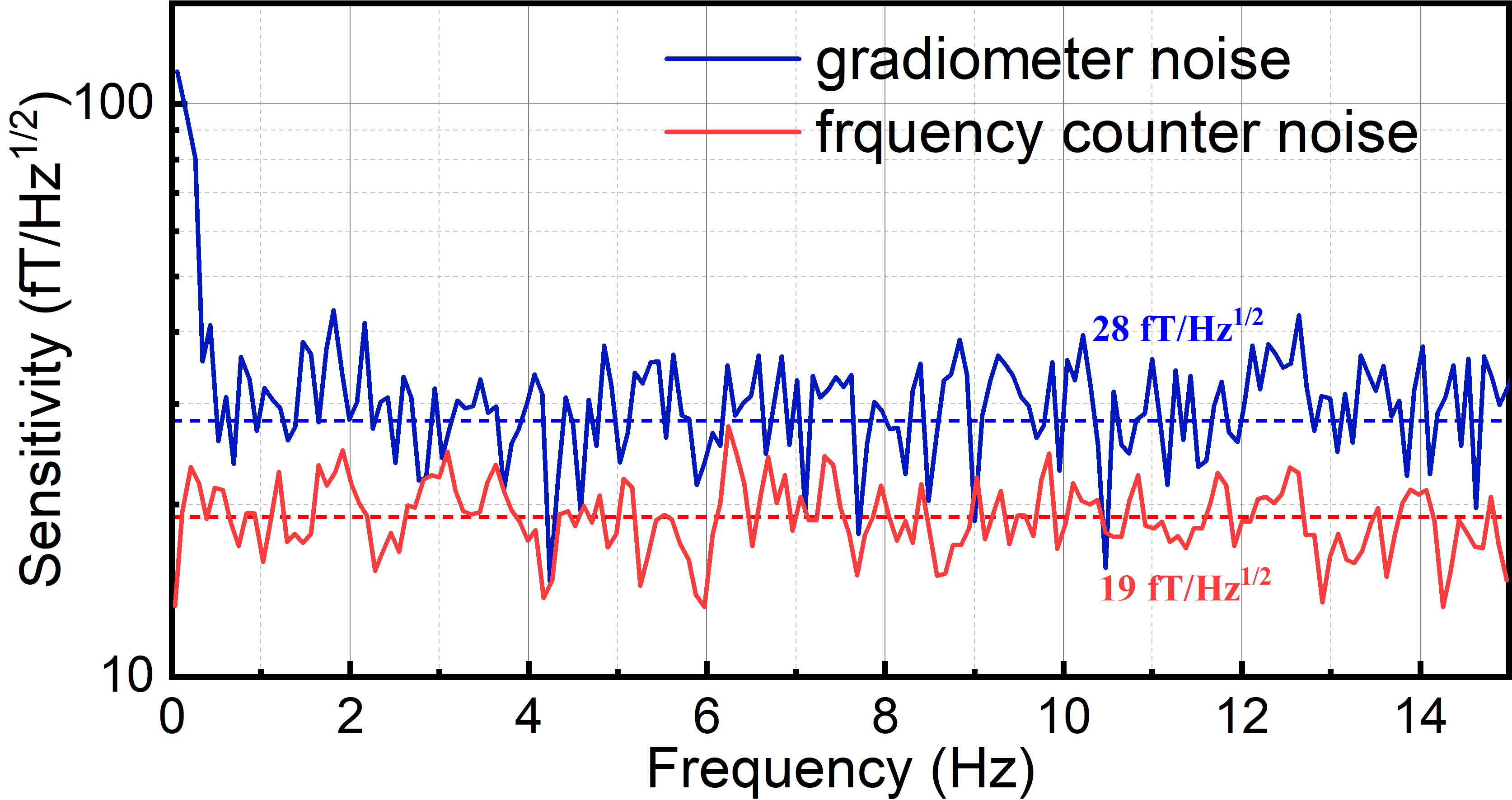}
  \caption{%
  Sensitivity measurement of the gradiometer. Blue: Amplitude spectrum of the magnetic-field gradient measured at $0.5~\mathrm{G}$. Red: Frequency-counter noise floor obtained from differential processing of the FID signal from a single vapor cell.
  }
  \label{fig:sensitivity}
\end{figure}

To study the intrinsic performance of the magnetometer, a gradiometric configuration is employed, where the differential frequency signal between the two channels is obtained through frequency-ratio measurements, which suppress common-mode fluctuations of the bias field. The magnetic-field differential noise spectrum is then extracted by converting frequency-ratio fluctuations into magnetic-field uncertainties using the known gyromagnetic ratio. At a bias field of $0.5~\mathrm{G}$, the gradiometer achieves a differential sensitivity of $28~\mathrm{fT}/\sqrt{\mathrm{Hz}}$ over the $0.5$--$15~\mathrm{Hz}$ bandwidth, as shown in Fig.~\ref{fig:sensitivity}. At bias fields of $0.3~\mathrm{G}$ and $0.7~\mathrm{G}$, the corresponding sensitivities are $32~\mathrm{fT}/\sqrt{\mathrm{Hz}}$ and $25~\mathrm{fT}/\sqrt{\mathrm{Hz}}$, respectively.

The noise contribution are further analyzed. We perform a reference measurement in which the signal from one channel is split and sent into both channels of the frequency counter. Because the two chancels receive the same signal, the differential output represents the counter's intrinsic noise, which is measured to be $19~\mathrm{fT}/\sqrt{\mathrm{Hz}}$. The spin-projection-noise and photon-shot-noise limits are estimated to be $6.5~\mathrm{fT}/\sqrt{\mathrm{Hz}}$ and $2.9~\mathrm{fT}/\sqrt{\mathrm{Hz}}$, respectively; the derivations are provided in Appendix~\ref{sec:noise_calculation}. The remaining excess noise is attributed to residual non-common-mode contributions, including mismatch between the two electronic channels of the frequency counter and environmental perturbations such as air convection and mechanical vibrations that affect the two optical paths differently.

\section{Conclusion}
In conclusion, we have demonstrated a paraffin-coated planar reflective multipass vapor cell for compact optical atomic magnetometry. The integration of high-reflection dielectric films on the inner walls of the vapor cell provides a mechanically robust multipass geometry without externally mounted mirrors, reducing optical loss and alignment sensitivity while retaining a simple planar structure compatible with anti-relaxation coatings. We show that paraffin coatings remain effective in this multipass architecture, providing longitudinal relaxation times exceeding $1~\mathrm{s}$ together with high optical transmission. Using two such cells, we implement an FID gradiometer and achieve a differential sensitivity of $\sim 28~\mathrm{fT}/\sqrt{\mathrm{Hz}}$ in the geomagnetic-field range. These results establish anti-relaxation-coated planar multipass vapor cells as a compact, low-temperature, and integration-friendly platform for ultrahigh-sensitive magnetometry.

\begin{acknowledgments}
This work was supported by National Science and Technology Major Project for Deep Earth Probe and Mineral Resources Exploration (2024ZD1002702).
\end{acknowledgments}

\appendix

\section{Noise calculation of the FID magnetometer}
\label{sec:noise_calculation}

In the FID magnetometer, the Larmor frequency is related to the magnetic field through
\begin{equation}
    \omega = \gamma B,
\end{equation}
where $\gamma$ is the gyromagnetic ratio. In our analysis, $B$ is estimated from the time between two zero crossings of the FID signal separated by $M$ oscillation cycles. Denoting the corresponding times by $t_i$ and $t_f$, with
\begin{equation}
    \Delta t \equiv t_f - t_i,
\end{equation}
the estimated values for $\omega$ and $B$ are
\begin{equation}
    \omega = \frac{2\pi M}{\Delta t},
    \qquad
    B= \frac{2\pi M}{\gamma \Delta t}.
    \label{eq:B_est_app}
\end{equation}
Assuming independent timing uncertainties at the two endpoints, the magnetic-field uncertainty is
\begin{equation}
    \delta B
    =
    \frac{2\pi M}{\gamma \Delta t^2}
    \sqrt{\delta t_i^2 + \delta t_f^2}
    =
    \frac{\omega}{\gamma \Delta t}
    \sqrt{\delta t_i^2 + \delta t_f^2}.
    \label{eq:dB}
\end{equation}

The FID signal can be approximated in the simplified form
\begin{equation}
    X(t) =
    X_0\exp\left(-\Gamma_{2}t\right)\sin(\omega t + \phi_0),
    \label{eq:Xt}
\end{equation}
where $\Gamma_2$ is the transverse relaxation rate. At a zero crossing point, $\sin(\omega t + \phi_0)=0$, so the local slope is
\begin{equation}
    \left|\frac{\partial X}{\partial t}\right|_{\mathrm{ZC}}
    =
    X_0 \omega\exp\!\left(-\Gamma_{2}t\right).
    \label{eq:slope}
\end{equation}
Therefore, the timing uncertainty of a single zero crossing is
\begin{equation}
    \delta t_k
    =
    \frac{\delta X_k}{X_0 \omega}\exp\left(\Gamma_{2}t_{k}\right),
    \label{eq:dt_local}
\end{equation}
with $\delta X_k$ the rms noise amplitude of the measured observable.
\subsection{Photon shot noise}

The photon shot noise is estimated from the optical power incident on a single detector channel. For a single-channel optical power $P$, the shot-noise-limited power fluctuation within bandwidth $\mathrm{BW}$ is
\begin{equation}
    \delta P = \sqrt{2P\epsilon \mathrm{BW}},
\end{equation}
where $\epsilon=h\nu$ is the photon energy. In a balanced polarimeter, the shot noise from the two detector channels adds in quadrature, giving
\begin{equation}
    \delta V = \sqrt{2}kG\delta P = 2kG\sqrt{P\epsilon \mathrm{BW}}.
    \label{eq:psn_voltage}
\end{equation}

The FID voltage is written as
\begin{equation}
    V(t) = V_0 \exp\left(-\Gamma_{2}t\right)\sin(\omega t+\phi_0),
\end{equation}
where $V_0$ is the initial FID amplitude. At a zero crossing, the voltage-to-time conversion gives
\begin{equation}
    \delta t_{k,\mathrm{psn}} = \frac{\delta V}{V_0\omega}\exp\left(\Gamma_{2}t_{k}\right).
\end{equation}
Substituting the timing uncertainties at $t_i$ and $t_f$ into Eq.~\eqref{eq:dB} yields
\begin{equation}
    \delta B_{\mathrm{psn}}
    =
    \frac{1}{\gamma\Delta t}\frac{\delta V}{V_0}
    \sqrt{\exp\left(2\Gamma_2 t_i\right)+\exp\left(2\Gamma_2 t_f\right)}
    \label{eq:dB_psn_general}
\end{equation}
For $t_i=0$ and $t_f=\Delta t$, using Eq.~\eqref{eq:psn_voltage}, this becomes
\begin{equation}
    \delta B_{\mathrm{psn}}
    =
    \frac{2kG}{\gamma\Delta t V_0}\sqrt{P\epsilon \mathrm{BW}
    (1+\exp\left(2\Gamma_2 \Delta t\right))}.
    \label{eq:dB_psn}
\end{equation}

For the numerical estimate, we use a Thorlabs P210A detector with a transimpedance conversion gain of $G=10~\mathrm{V/mW}$ and an RF amplification factor of $k=27.8$. The other parameters are $\gamma/2\pi=7~\mathrm{Hz/nT}$, $\Gamma_2\approx 33~\mathrm{s^{-1}}$, $\Delta t=27.5~\mathrm{ms}$, $\mathrm{BW}=15~\mathrm{Hz}$, $P=140~\mu\mathrm{W}$, and $V_0=10~\mathrm{V}$. For the FID gradiometer, the noise from the two channels is assumed to be independent, so the differential noise is larger than the single-channel value by a factor of $\sqrt{2}$. Equation~\eqref{eq:dB_psn} then gives a photon-shot-noise-limited sensitivity of $2.9~\mathrm{fT}/\sqrt{\mathrm{Hz}}$ for the gradiometer.

\subsection{Spin-projection noise}

Consider an ensemble of $N$ atoms within the probe beam volume, with total angular momentum $F$. The atoms are optically pumped into a stretched state along $z$ and tipped by an rf pulse through an angle $\theta$. The mean collective spin projection measured along $x$ is

\begin{equation}
    \langle J_x(t) \rangle = N F \sin\theta \exp\left(-\Gamma_{2}t\right)\sin(\omega t + \phi_0),
\end{equation}
so that the zero-crossing slope is
\begin{equation}
    \left|\frac{\partial \langle J_x \rangle}{\partial t}\right|_{\mathrm{zc}} = N F \sin\theta\,\omega \exp\left(-\Gamma_{2}t\right).
    \label{eq:Jx_slope}
\end{equation}

For a coherent spin state, the projection noise along an axis perpendicular to the spin polarization direction is
\begin{equation}
    \Delta J_\perp^2 = \frac{N F}{2}.
\end{equation}
Thus, at zero crossings where the probe axis is perpendicular to the mean spin, the corresponding timing uncertainty is
\begin{equation}
    \delta t_{k,\mathrm{spn}} = \frac{1}{\omega \sin\theta \sqrt{2 N F}}\exp\left(\Gamma_{2}t_{k}\right).
    \label{eq:dt_spn}
\end{equation}
Owing to the long longitudinal relaxation time provided by the anti-relaxation coating, longitudinal relaxation during the probing interval is neglected. Because the probe beam area is much smaller than the cell cross section, most atoms observed at $t_f$ are not the same atoms observed at $t_i$. Therefore, the correlations between the projection noise at $t_i$ and $t_f$ are negligible~\cite{Tang2020}, and the two noise contributions are treated as statistically independent. The corresponding magnetic-field uncertainty
for $t_i = 0$ and $t_f = \Delta t$ is then
\begin{equation}
    \delta B_{\mathrm{spn}} = \frac{\csc\theta}{\gamma \Delta t \sqrt{2 N F}}\sqrt{1+\exp\left(2\Gamma_2 \Delta t\right)}.
    \label{eq:dB_spn}
\end{equation}
In our experiment, the spin tilt angle is $\theta = \pi/12$ and the angular momentum is $F = 2$.  The atomic number density of rubidium atoms in the vapor cell is estimated from the saturated vapor pressure to be $n = 1.5 \times 10^{11}~\mathrm{cm^{-3}}$~\cite{Steck2001}. The volume probed by the detection beam is $V_p = 0.19~\mathrm{cm^3}$, giving $N = nV_p = 2.85 \times 10^{10}$. The other parameters, including $\gamma$, $\Gamma_2$, and $\Delta t$, are the same as those used in the photon-shot-noise estimate. Substituting them into Eq.~\eqref{eq:dB_spn} yields a spin-projection-noise-limited sensitivity of $6.7~\mathrm{fT}/\sqrt{\mathrm{Hz}}$.
\bibliography{references}
\end{document}